\documentstyle[11pt,aaspp4,psfig]{article}

\received{}
\accepted{}
\journalid{}{}
\articleid{}{}

\slugcomment{ to appear in \apj}

\lefthead{Krennrich et al.}
\righthead{Detection of Multi-TeV Emission from Markarian 421}

\begin{document}

\def\lsim{\mathrel{\hbox{\rlap{\hbox{\lower4pt\hbox{$\sim$}}}\hbox{$<$}}}}
\def\gsim{\mathrel{\hbox{\rlap{\hbox{\lower4pt\hbox{$\sim$}}}\hbox{$>$}}}}

\title{Gamma-ray Observations of the Galactic Plane at\\ 
Energies E $>$ 500 GeV}

\author{S.LeBohec\altaffilmark{1},
I.H.Bond \altaffilmark{2}, 
S.M.Bradbury \altaffilmark{2},
J.H.Buckley\altaffilmark{3},
A.M.Burdett\altaffilmark{2,4}, 
D.A.Carter-Lewis\altaffilmark{1}, 
M.Catanese\altaffilmark{4}, 
M.F.Cawley\altaffilmark{5}, 
S.Dunlea\altaffilmark{6}, 
M. D'Vali\altaffilmark{2}, 
D.J.Fegan\altaffilmark{6}, 
S.J.Fegan\altaffilmark{4}, 
J.P.Finley\altaffilmark{7}, 
J.A.Gaidos\altaffilmark{7}, 
T.A.Hall\altaffilmark{7}, 
A.M.Hillas\altaffilmark{2}, 
D.Horan\altaffilmark{4,6}, 
J.Knapp\altaffilmark{2}, 
F.Krennrich\altaffilmark{1},
R.W.Lessard\altaffilmark{7}, 
D. Macomb\altaffilmark{8},
C.Masterson\altaffilmark{6}, 
J.Quinn\altaffilmark{6}, 
H.J.Rose\altaffilmark{2}, 
F.W.Samuelson\altaffilmark{1},
G.H.Sembroski\altaffilmark{7}, 
V.V.Vassiliev\altaffilmark{4},
T.C.Weekes\altaffilmark{4}}

\begin{abstract}

In 1998 and 1999 the Whipple Observatory 10 m telescope was used to
search for diffuse gamma ray emission from the Galactic Plane.  At this
time, the telescope was equipped with a large ({$\rm4.8^o$}) field of
view camera, well suited to detect diffuse $\gamma$-ray emission. No
signifiant evidence of emission was found.
Assuming the TeV emission profile matches EGRET observations above 1~GeV
with a differential spectral index of $2.4$, we derive an upper
limit of { $\rm {3.0\cdot10^{-8}\:cm^{-2}\:s^{-1}\:sr^{-1}}$} for the
average diffuse emission above {{$\rm500~GeV$}} in the galactic 
latitude range from {{$\rm-2^o$ to $\rm +2^o$}} at galactic 
longitude {{$\rm 40^o$}}. Comparisons with EGRET observations
provide a lower limit of 2.31 for the differential spectral index of
the diffuse emission, assuming there is no break in the spectrum
between 30~GeV and 500~GeV. This constrains models for diffuse
emission with a significant inverse Compton contribution.  
\end{abstract}


%C. W. Akerlof\altaffilmark{2},
%S. D. Biller\altaffilmark{3},
%J. H. Buckley\altaffilmark{4},
%A. M. Burdett\altaffilmark{3},
%D. A. Carter-Lewis\altaffilmark{1},
%M. F. Cawley\altaffilmark{5},
%M. Catanese\altaffilmark{1},
%V. Connaughton\altaffilmark{6},
%D. J. Fegan\altaffilmark{6},
%J. P. Finley\altaffilmark{7},
%J. A. Gaidos\altaffilmark{7},
%F. Krennrich\altaffilmark{1},
%R. C. Lamb\altaffilmark{1},
%R. Lessard\altaffilmark{6},
%J. E. McEnery\altaffilmark{6},
%D. Macomb\altaffilmark{8},
%G. Mohanty\altaffilmark{1},
%J. Quinn\altaffilmark{6},
%A. J. Rodgers\altaffilmark{3},
%H.J. Rose\altaffilmark{3},
%M. S. Schubnell\altaffilmark{2},
%G. H. Sembroski\altaffilmark{7},
%T. C. Weekes\altaffilmark{4},
%C. Wilson\altaffilmark{7}
%and J. Zweerink\altaffilmark{1}

\altaffiltext{1}{Department of Physics and Astronomy, Iowa State
University, Ames, IA 50011-3160}

%\altaffiltext{2}{Randall Laboratory of Physics, University of
%Michigan, Ann Arbor, MI 48109-1120}

\altaffiltext{2}{Department of Physics, University of Leeds,
Leeds, LS2 9JT, Yorkshire, England, UK}

\altaffiltext{3}{Department of Physics, Washington University, St. Louis, 
MO 63130, USA}

\altaffiltext{4}{ Fred Lawrence Whipple Observatory, 
Harvard-Smithsonian CfA, P.O. Box 97, Amado, AZ 85645-0097} 

\altaffiltext{5}{Physics Department, St.Patrick's College,
Maynooth, County Kildare, Ireland}

\altaffiltext{6}{Physics Department, University College, Dublin
4, Ireland}

\altaffiltext{7}{Department of Physics, Purdue University, West
Lafayette, IN 47907}

\altaffiltext{8}{Laboratory for High-Energy Astrophysics, 
NASA/GSFC, Greenbelt, MD 20771
and Universities Space Research Association, Lanham MD}

\clearpage

\section{Introduction}

High energy $\gamma$-rays traverse the Galaxy without significant
attenuation, and hence diffuse $\gamma$-ray emission probes the galaxy
as a whole.  As was observed with OSO-3 \cite{krau72}, emission
from the galactic plane is the main feature of the high energy
$\gamma$-ray sky.  Observations with the SAS-2 and COS-B satellites
showed this emission was generally correlated with the spatial structures
in the Galaxy seen at other wavelengths (\cite{fich75}; \cite{hart79};
\cite{maye80}). The greater sensitivity and angular resolution of the
EGRET instrument (\cite{hugh80}; \cite{camb88}; \cite{camb89})
provided a much more detailed spectral and spatial picture of the
diffuse emission from the Galaxy (\cite{hunt97}).

The spectral and spatial distributions of diffuse high energy emission
can be interpreted using a detailed model (\cite{bert93}) of the 
emission. This model is based on $\gamma$-ray
production by nucleon-nucleon and electron (bremsstrahlung and inverse
Compton) scattering in the interstellar medium.  The matter and photon
distributions in the interstellar medium are inferred from radio,
infrared, visible and ultra-violet observations.  The cosmic rays are
assumed to have the same composition and spectrum everywhere in the
galaxy with a density that is correlated with the matter density.  The
model precisely reproduces the spatial and spectral data from EGRET up
to energies of 100 MeV. However, for $\rm E>1~GeV$ the predicted flux
is 40\% lower than observations indicate (\cite{hunt97}).

\cite{pohl98} explained the observed excess at $\rm E>1~GeV$ as inverse
Compton (IC) emission by a cosmic-ray electron injection spectrum
harder than that used by Hunter et al.(1997). The latter assumed a spectral
index of 2.4, which was inferred from the local cosmic-ray electron
spectrum when propagated through the galaxy. If shock acceleration in 
supernova remnants (SNRs)
produces the electrons, a spectral index closer to 2.0 is expected.
Furthermore, most SNRs have power law spectra at radio wavelengths
with $\alpha \sim 0.5$ (\cite{gree95}), indicating injection spectra
with index near 2.0.  \cite{pohl98} showed that a
change of spectral index of only 0.4 is
sufficient to account for the observed excess.  Indeed, it has been
previously argued that if the electron cutoff energy is high enough,
IC emission may be the dominant source of diffuse $\gamma$-rays at TeV
energies (\cite{port97}). \cite{pohl98} suggest that the discrepancy
with the local electron spectrum is a statistical fluctuation: high
energy cosmic-ray electrons tend to have a local origin because
bremsstrahlung and IC energy losses prevent propagation over large 
distance. Therefore, they are subject to
Poisson fluctuations in the number of supernovae accelerating
electrons at a given time, whereas the unattenuated $\gamma$-rays
originate from broader parts of the Galaxy and reflect an electron
spectral index closer to the galactic average. 
In examining possible mechanisms for the EGRET excess, \cite{mos00} and Strong,
Moskalenko and Reimer (1999, 2000) apply several detailed models to
the full spectrum of diffuse gamma-ray emission from the Galaxy.  
These authors also conclude that IC
emission from a hardened electron injection spectrum ($<2$) is the
favored origin of the high energy EGRET excess.

We have searched for diffuse emission from the Galactic Plane at
higher energies than detected by EGRET using the Whipple Observatory 
10 m imaging atmospheric Cherenkov telescope to observe the galactic
plane region at ($b=0^o$, $l=40^o$).  Although a
simple extrapolation of the highest energy EGRET spectrum
indicates the diffuse emission should be detectable, no signal was 
found. Using an analysis method specifically developed for this 
measurement we
derive upper limits on the galactic diffuse $\gamma$-ray emission at
$\rm E>500~GeV$. We compare this result to a power law extrapolation of
EGRET's spectral points above 1~GeV and derive a lower limit on the
spectral index.  In sections 2 and 3 the observations and data
analysis are described. In section 4, the results are presented and
discussed in the context of the EGRET results summarized above.

\section{Observations}

The observations were made with the Whipple Observatory atmospheric Cherenkov
imaging telescope as described in \cite{caw90}.  The telescope
is located at an altitude of 2300m and consists of a 10 m
diameter optical reflector with a fast photomultiplier tube (PMT)
camera in the focal plane.  For point sources, the image shapes and
orientations can be used to distinguish $\gamma$-ray images from a much
larger number of cosmic-ray hadron images (\cite{hil85}).
The camera can also be used to map an extended portion of the sky
(\cite{less97}).  The observations
presented here were obtained using a 331 PMT camera with spacing of
0.24$^o$ covering a field of view of 4.8$^o$.

The observations of the galactic plane were centered on coordinates
($l=40^o ,b=0^o$), which is within the inner region of the galaxy
where the diffuse emission is most intense. This region culminates at
the relatively small zenith angle 25$^o$ at the Observatory 
and so gives a fairly low energy threshold.  The field was also 
selected for its lack of bright stars
which allows sensitive atmospheric Cherenkov radiation measurement.

The observations were carried out in a standard ON-OFF mode in which
the galactic plane (ON) is tracked for 28 minutes after which the
telescope tracks a background region (OFF) covering the same path in
elevation and azimuth for another 28 minutes.  Some of the OFF data were
measured before the ON data.  During the observations, several
PMTs were switched off in both ON and OFF data set because of light 
from bright stars. 

In 1998, the telescope was triggered when any two out of the 331 PMTs
exceeded a threshold corresponding to $\sim65$ photoelectrons.
From Monte Carlo simulations we found that, if analyzed as describe
below, the 1998 data had an energy threshold of 700~GeV. In 1999 a 
topological (pattern) trigger was used requiring triggering PMTs to 
be neighbors (\cite{brad99}), and we  
reduced the threshold to 500~GeV.  (Here, the telescope energy 
threshold is defined
as the energy for which the count rate peaks for a power-law spectrum
with differential index 2.4). The data consists of 7 ON-OFF pairs obtained
in 1998 and 10 ON-OFF pairs in 1999.
  
\section{Data analysis}

In this section we describe the main four steps in the analysis, starting
with the $\rm \gamma$-ray candidate selection (step~1) and angular
origin reconstruction (step~2). For this particular analysis we also
had to correct for sensitivity inhomogeneities across the field (step~3). 
This allowed us to fit a specific pattern corresponding to diffuse
emission in the final data (step~4).

\subsection{Gamma-ray candidate selection}

Software ``padding'' (\cite{caw93}) is used to compensate for
differences in noise level in the ON and OFF runs. The image is also
``cleaned'' (\cite{rey93}) to alleviate the noise effects of pixels that
are not
part of the shower image. The image is then characterized by the image
parameters calculated from the $1^{st}$, $2^{nd}$ and
$3^{rd}$ moments of the recorded light distribution(see Figure 1). The
image parameters are {\it Length}, {\it Width}, {\it
Distance} from the center of the field of view and {\it
Asymmetry} as described in \cite{feg97}.  The $\gamma$-ray images tend
to be comet-shaped, with {\it Asymmetry} indicating the
development direction of the shower. At least 78 and 56 photoelectrons
were required in the two brightest pixels of each image to ensure that
it is well above background noise. The {\it Distance} of the image
centroid from the center of the field of view was restricted to be
less than $\rm 1.8^o$ to ensure that the entire image is contained in the
camera. Boundary values for the {\it Length} and {\it Width}
parameters were derived by maximizing the significance of the Crab
Nebula observations obtained with this camera. These
selection criteria for $\gamma$-ray candidates are summarized in Table
1. Since the telescope is being used to image an extended portion of
the sky, no orientation selection is made.

\subsection{Reconstruction of the arrival direction}

For each selected $\gamma$-ray candidate, the most likely point of
origin is reconstructed (see Figure 1) by applying the following
method (\cite{less97}) which was tested and optimized on off-center
Crab Nebula observations:\\
%\begin{center}
$\bullet$ the source is assumed to lie on the shower image major axis,\\
$\bullet$ the source is assumed to be in the direction indicated by the 
$Asymmetry$ parameter, and\\
$\bullet$ the source is assumed to be located $ 1.85^o(1-Width/Length)$ 
from the image centroid.
%\end{center}

The first criterion results from the fact that the image axis is the
projection in the focal plane of the shower axis which points toward
the source.  The $Asymmetry$ parameter allows the identification of
the top and the bottom of the shower. The third criteria is empirical
but intuitive. The image elongation (measured by $1-Width/Length$) and
its distance from the source both increase with the shower impact parameter.
The standard deviation of the point spread function provided by this
method is $\rm 0.12^o$.

\subsection{Sensitivity correction across the field}

As one moves from the center of the field of view, the sensitivity
decreases because the outermost portions of the images may be
missed. In addition, since pixels were turned off because of stars,
the sensitivity profile has irregularities.  We have developed the
following calibration procedure to compensate for the variation of
sensitivity across the field of view.

A uniform distribution of TeV {$\gamma$}-rays was simulated. The
selection criteria from section 3.1 were applied. Figure 2 shows the radial
distribution of the reconstructed source position which is well fit by
a centered two-dimensional Gaussian with a standard deviation of
{$1.2^o$}. We have verified this sensitivity pattern at two points using
Crab Nebula data taken off axis. We then used background events that 
pass our selection criteria to measure $S_{correct}$, the discrepancy 
ratio between the simulated and real field sensitivity functions.  
The $S_{correct}$ factor map obtained for the 1999 galactic plane 
observations is shown
in Figure 3.  Correction factors can be as large as 2. In the
subsequent analysis, each event has a weight equal to
the $1/S_{correct}$ factor corresponding to the reconstructed
direction of origin in the camera frame.  This correction allows us to
directly compare the simulated diffuse emission with the recorded
data.

\subsection{Galactic diffuse emission flux measurement}

In this analysis, we assume that the diffuse emission is uniform
along the galactic equator over our field of view. The latitude
profile has been taken from EGRET observations above 1~GeV as shown in
Figure 4, where the smooth curve indicates how the profile was
modeled. We have used simulations to estimate the response of the camera to
this emission profile assuming a power law energy spectrum with a
differential spectral index of 2.4.  As one moves across the camera in
the direction of galactic longitude, the number of {$\gamma$}-rays
detected should remain constant for an ideal camera.  However, because
of the finite field of view, the detection efficiency falls off toward
the edge. Consequently, we have restricted the data to longitudes
between {$\rm 38.5^o$ and $\rm 41.5^o$} (left side of Figure 5).  
The resulting latitude profile as it should appear in our data is 
shown on the right side of Figure 5. The smooth curve shows how it 
was modeled when fit to our data with amplitude as the only free parameter.

\section{Results}

The analysis described above was applied to the data accumulated in
1998 and 1999. The latitude profile derived from the simulation of the
diffuse emission was fitted to both the 1998 and 1999 data with
results shown in Figure~6. Although the 1998 data exhibited a
$3.2\sigma$ excess ({$\rm (1.84^+_-0.57)\gamma/minute$}), the 1999 data is
consistent with a null result ($\rm (0.42^+_-0.43)\gamma/minute$), and,
from both datasets we derive upper limits on the average diffuse
emission flux in the portion of sky defined by 
{$\rm 38.5^o<l<41.5^o$} and {$\rm 2^o<b<2^o$}.  
Assuming a differential 
spectral index $\gamma=2.4$, we obtained a {$99.9\% $} confidence
level upper limit on the flux of 
$\rm 6.3\cdot 10^{-8}s^{-1}cm^{-2}sr^{-1}$ for 1998 with a
threshold of 700~GeV and $\rm 3.0\cdot 10^{-8}s^{-1}cm^{-2}sr^{-1}$ for
1999 with a threshold of 500~GeV. 

In Figure 7, simple power law spectra corresponding to the upper
limits using several values of spectral index are compared to the high
end of the spectrum obtained with EGRET for the same portion of sky.
Our results are not compatible with an extrapolation of the EGRET
measurements assuming a hard spectrum. This is quantified in Figure 8
which shows the {$99.9\% $} and {$99\% $} confidence level upper
limits on the diffuse emission flux above 500~GeV as a function of the
assumed spectral index. It is compared to the simple power law
extrapolation to $\rm E>500~GeV$ of the EGRET measurement above 1~GeV
for the same field. This comparison leads to a lower limit on the
differential spectral index of 2.31 at the $99.9\%$ confidence level.

\section{Concluding Remarks}

The observations of the galactic plane did not yield a detection of 
the diffuse emission, and we derived an upper limit of $\rm
3.0\cdot10^{-8}s^{-1}cm^{-2}sr^{-1}$ for the flux above 500~GeV in the
portion of galactic plane defined by {$\rm 38.5^o<l<41.5^o$} and {$\rm
2^o<b<2^o$}, assuming a differential spectral index of 2.4. We
showed that these upper limits imply that the diffuse emission cannot
extend up to 500~GeV with a simple power law with a differential
spectral index smaller than 2.31.

These observations are compatible with earlier observations made in the
drift-scan mode with the 10m reflector (\cite{rey93}; \cite{rey90}). 
Other constraints on the
extrapolation of the diffuse emission spectrum to greater energies
than observed by EGRET, arise from observations with extensive air
shower detectors like TIBET above 10~TeV (\cite{ameno97}), EAS-TOP
(\cite{agli92}) and  CASA MIA (\cite{bori98}) above 130~TeV. Among these
experiments, TIBET, the closest one to us on the energy scale,
also concentrated on the portion of sky defined by {$\rm
20^o<l<55^o$} and {$\rm -5^o<b<+5^o$} including the field we observed.
These observations provided similar limits on the spectral index 
as presented here but the energies being probed are larger by at least
one order of magnitude.
 
In the Hunter et al. (1997) model, a value of $\sim 2.3$ for the
inverse Compton spectrum for the upper end (50~GeV) of the EGRET
spectrum is found from Fig.~4 of their paper.  This model assumes a
SNR electron injection spectral index of 2.4, and the inverse Compton
gamma-ray spectral index is marginally consistent with our lower
limit.  However, the model does not fit the high end of the EGRET
spectrum.  \cite{pohl98} explain the excess
above 1~GeV as due to inverse Compton emission in a model using
electrons injected into the galaxy from SNR with a harder power-law
spectrum with an index of 2.  However, their inverse 
Compton spectrum gives a differential spectral index of $\sim 1.85$, 
in contradiction our limit of 2.31.
In extrapolating to energies 
of 500~GeV, there may be some softening of the inverse Compton spectrum 
due to the onset of Klein-Nishina reduction in scattering cross section.
Nevertheless  the Cosmic Background Radiation should be a 
large part of the target 
photon population, and the corresponding portion of the inverse Compton
spectrum would not be significantly affected until the
$\gamma$-rays reach energies well above 500~GeV.  Thus our limits
should constrain such models (e.g., Pohl and Esposito,
1998; Strong, Moskalenko and Reimer, 2000) possibly implying a break in 
the primary electron spectrum.

\acknowledgments

We acknowledge the technical assistance of K.~Harris and
E.~Roach. This research is supported by grants from the U.S.~Department
of Energy and by PPARC in the UK and by Forbairt in Ireland.

\vfill\eject

\clearpage

 \clearpage
 \begin{deluxetable}{lrrrrrr}
 \footnotesize
 \tablecaption{$\gamma$-ray candidate selection.
 \label{tbl-1}}
 \tablewidth{0pt}
 \tablehead{
 \colhead{} &
% \colhead{} & 
% \colhead{} &
 \colhead{}  & \\
 \colhead{Selection criteria} &
% \colhead{}
}
 \startdata
 $Max1\rm>78\,d.c.$ &\nl 
 $Max2\rm>56\,d.c.$ & \nl 
 $\rm0.073^o<\it Width\rm<0.16^o$ & \nl 
 $\rm0.16^o<\it Length\rm<0.43^o$ & \nl
 $Distance\rm<1.8^o$ &\nl
\enddata
 \end{deluxetable}

\vfill\eject

\clearpage
\begin{center}
\mbox{\epsfysize=0.60\textheight\epsfbox{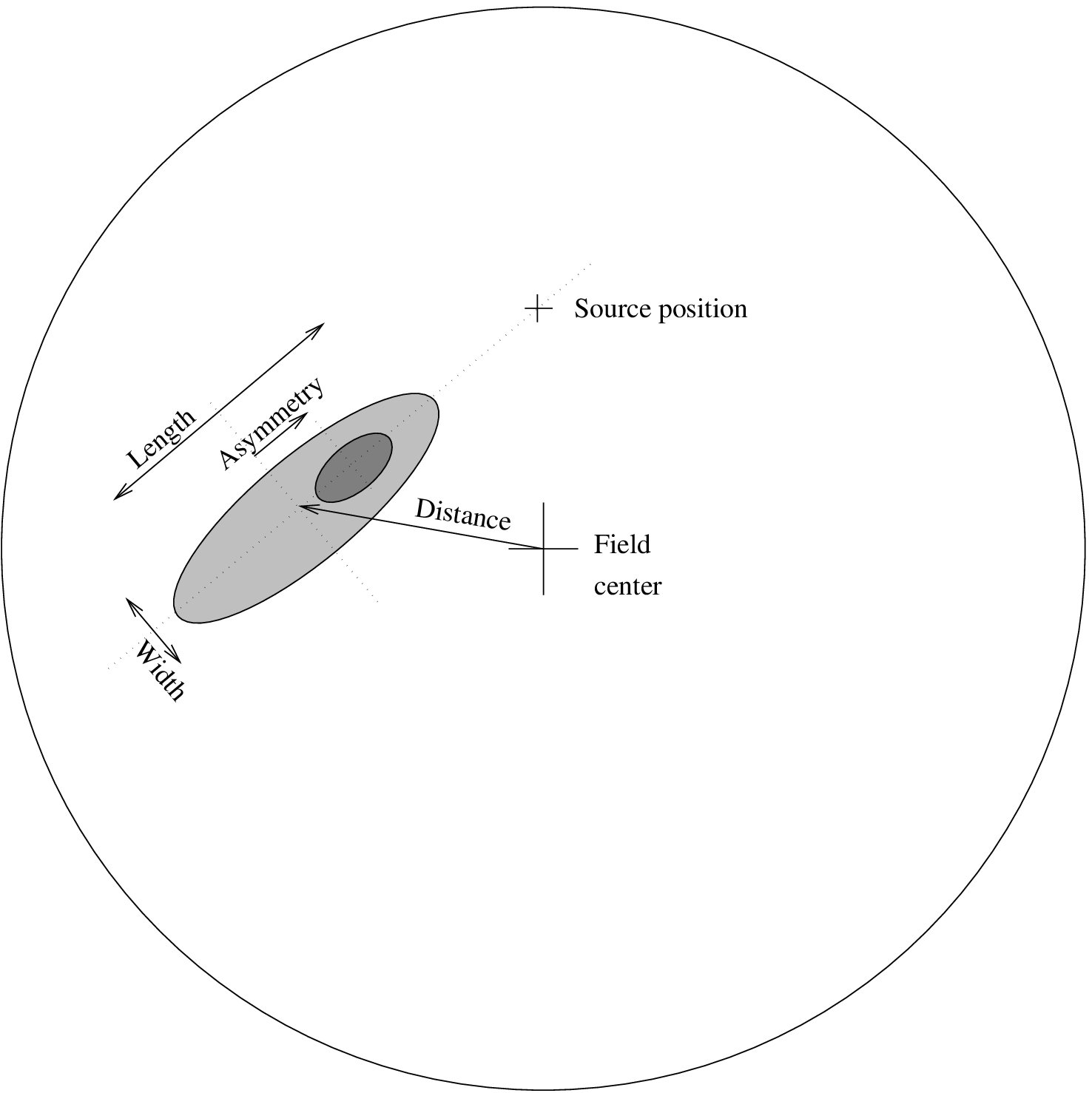}}
\end{center}
 \figcaption[m42d]
 {The elliptical image is first characterized by its position and
orientation. The $Length$ and  $Width$ allow the preferential
selection of $\gamma$-ray shower images which tend to be more
elongated than the bulk of cosmic-ray shower images. The asymmetry 
in the distribution of light along the major
axis can be characterized by the $Asymmetry$ parameter and indicates 
the direction of the top of the shower and therefore points to the 
source direction. 
 \label{fig1}}

\clearpage
\begin{center}
\mbox{\epsfysize=0.60\textheight\epsfbox{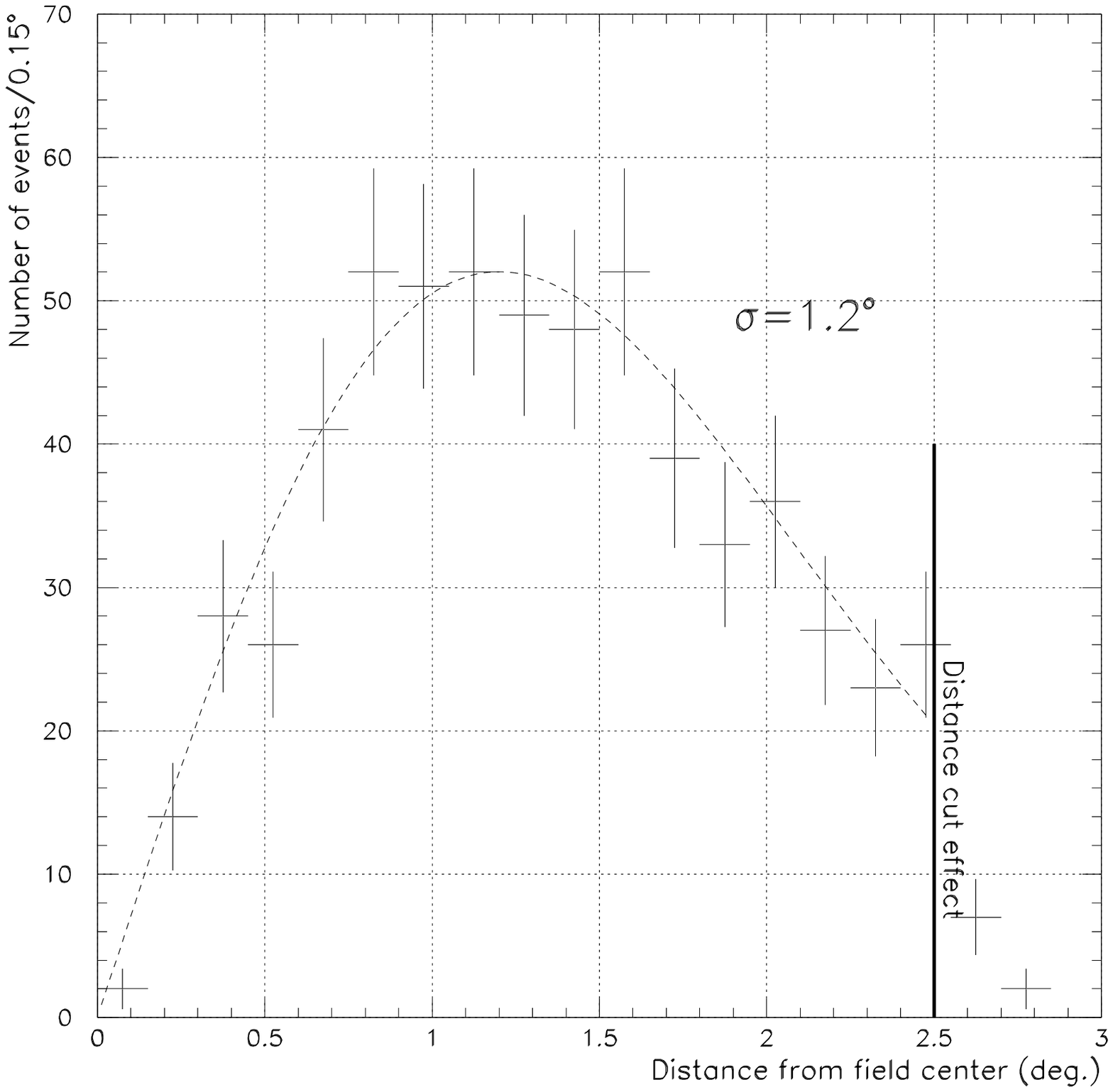}}
\end{center}
 \figcaption[m42d] {Radial distribution of the reconstructed arrival
directions of simulated $\gamma$-rays with homogeneous arrival
direction. The distribution is compared with the radial distribution
of a two dimensional Gaussian with a standard deviation of
$1.2^o$. The image centroid $distance$ cut at $1.8^o$ prevents 
reconstruction of events further than $2.5^o$ away from the field
center.
 \label{fig2}}

\clearpage
\begin{center}
\mbox{\epsfysize=0.60\textheight\epsfbox{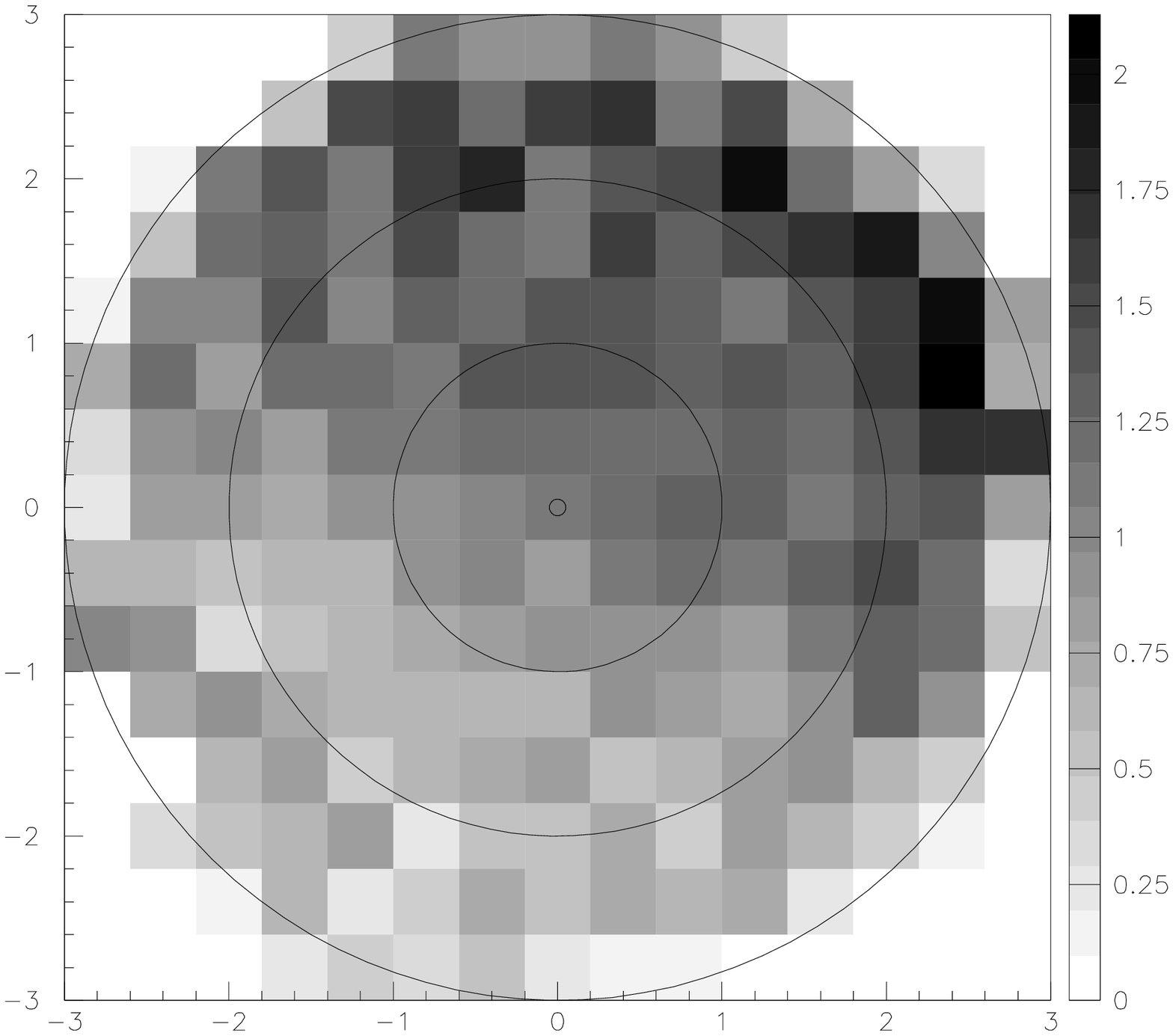}}
\end{center}
 \figcaption[m42d]
{The sensitivity correction factor $S_{correct}$ is the ratio between the
sensitivity inferred from simulation and the sensitivity to $\gamma$-ray
like background events. This figure represent the camera frame map of 
$S_{correct}$ obtained from the 1999 OFF-source galactic plane observation. 
 \label{fig3}}

\clearpage
\begin{center}
\mbox{\epsfysize=0.60\textheight\epsfbox{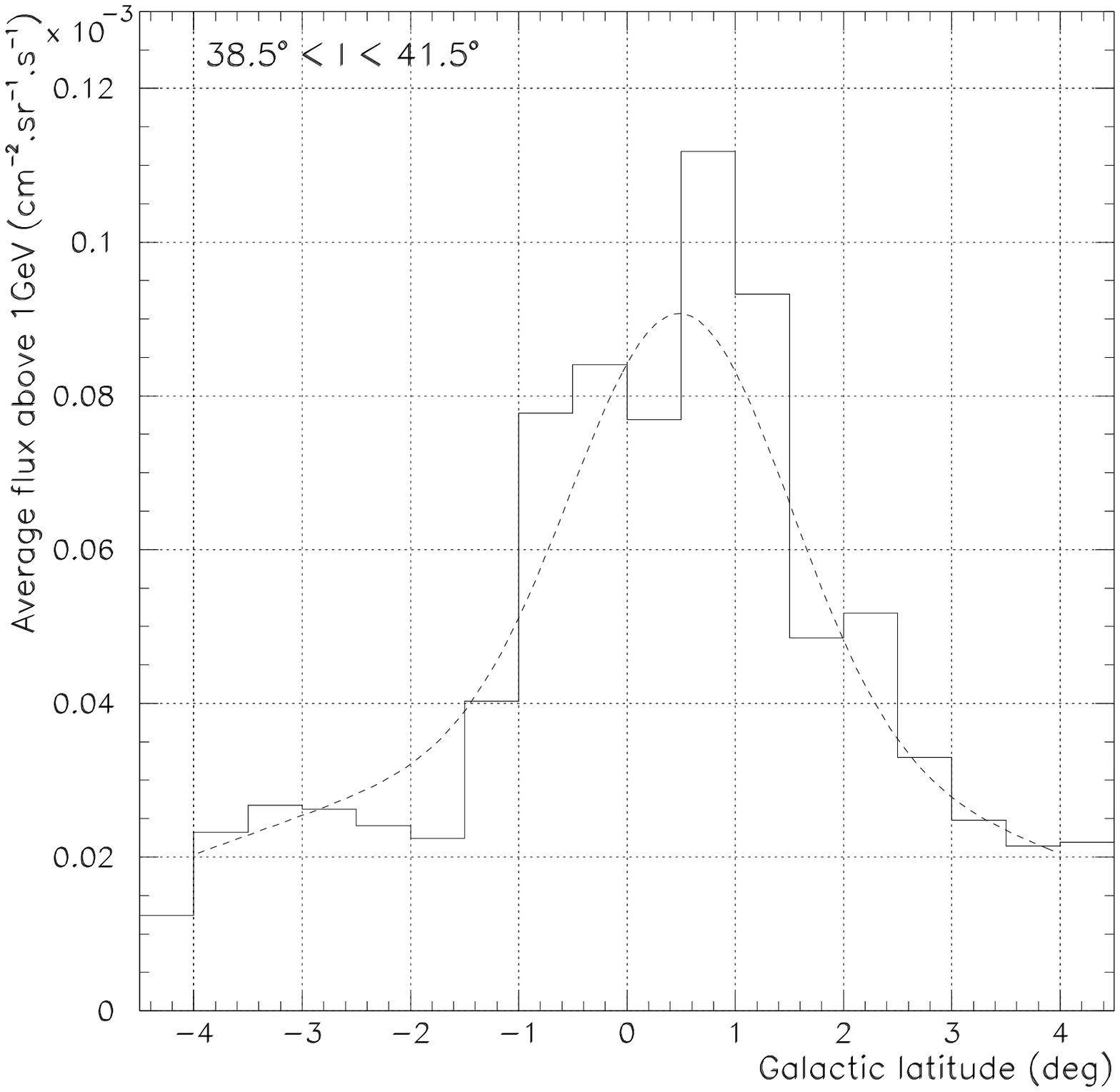}}
\end{center}
 \figcaption[m42d] {Average diffuse
emission latitude profile between the galactic longitude $l=38.5^o$
and $l=41.5^o$ as observed by EGRET at energies $\rm E>1~GeV$. The
smooth curve indicates how the distribution was modeled for the
diffuse emission simulation above 500GeV.
 \label{fig4}}

\clearpage
\begin{center}
\mbox{\epsfysize=0.40\textheight\epsfbox{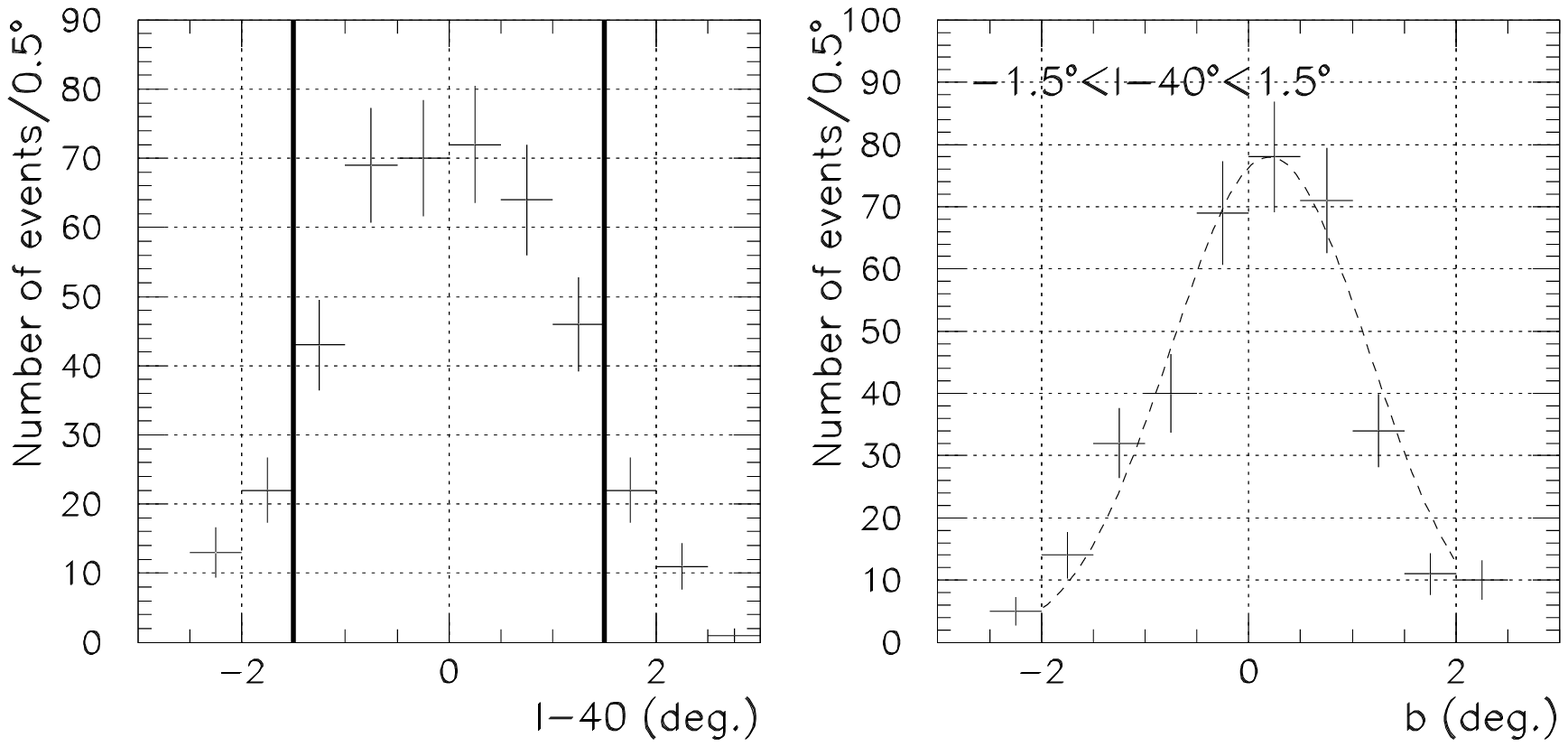}}
\end{center}
 \figcaption[m42d] {Distribution of the reconstructed direction of
origin for the simulated TeV diffuse emission when projected on the
galactic equator (left) and on the meridian (right). This distribution
is obtained after all the selection criteria have been applied. The
two vertical lines on the left histogram indicate the longitude
selection criteria.
 \label{fig5}}

\clearpage
\begin{center}
\mbox{\epsfysize=0.30\textheight\epsfbox{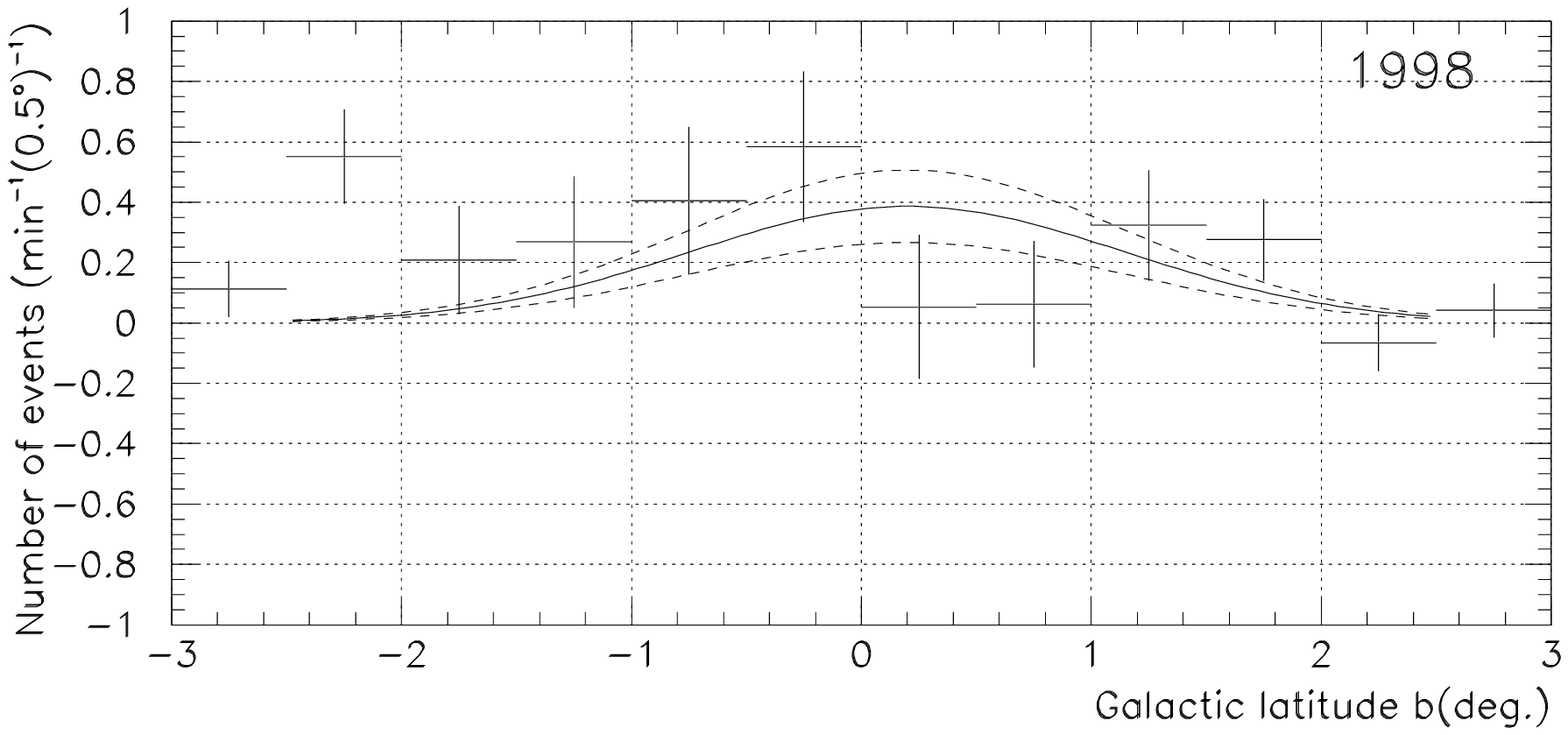}}
\mbox{\epsfysize=0.30\textheight\epsfbox{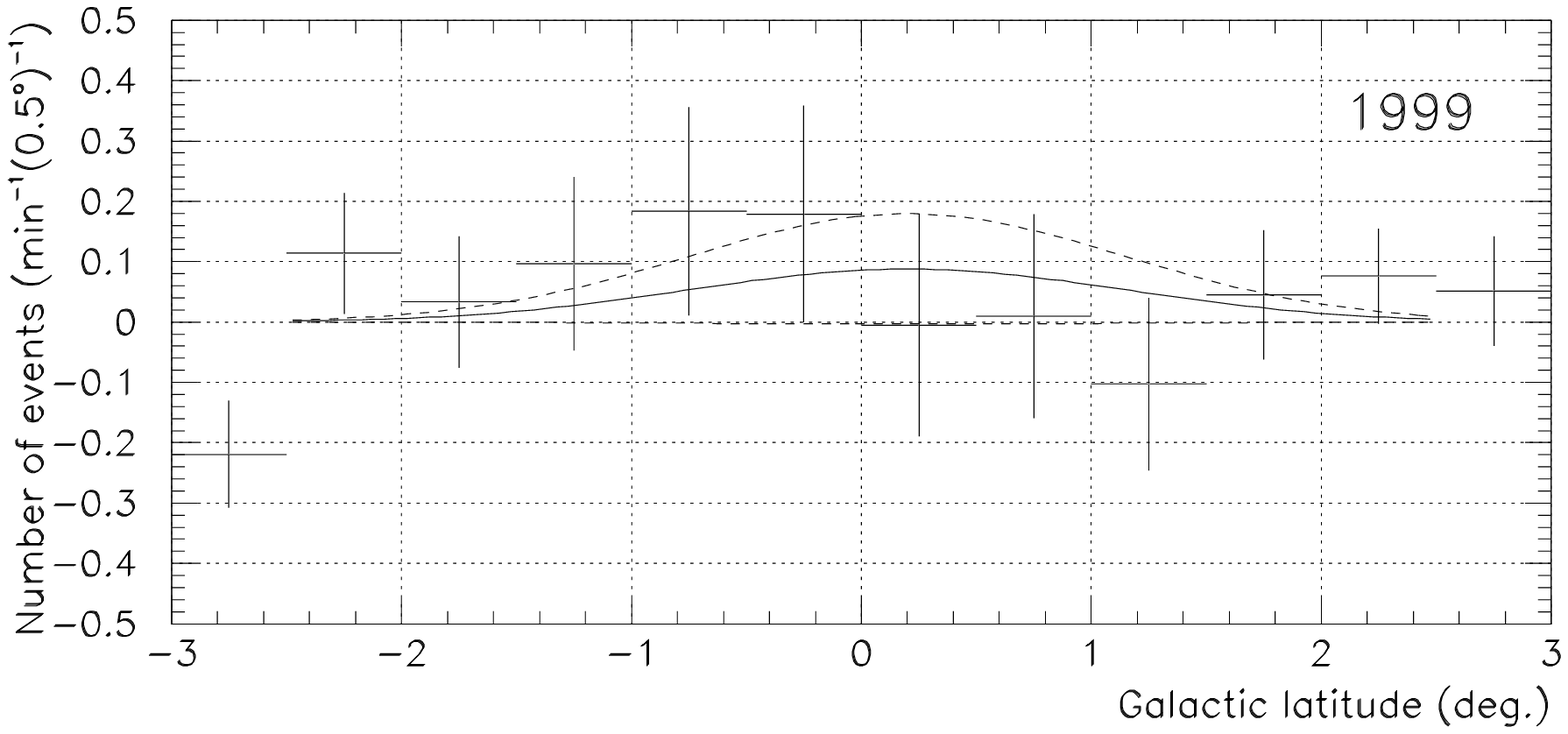}}
\end{center}
 \figcaption[m42d] {Latitude distribution of the excess observed in
1998 (top) and 1999 (bottom). The smooth solid curve shows the
modeled simulated latitude profile when fitted to the data with
amplitude as the only free parameter. The smooth dashed lines
correspond to one standard deviation about the best fit.
 \label{fig6}}

\clearpage
\begin{center}
\mbox{\epsfysize=0.60\textheight\epsfbox{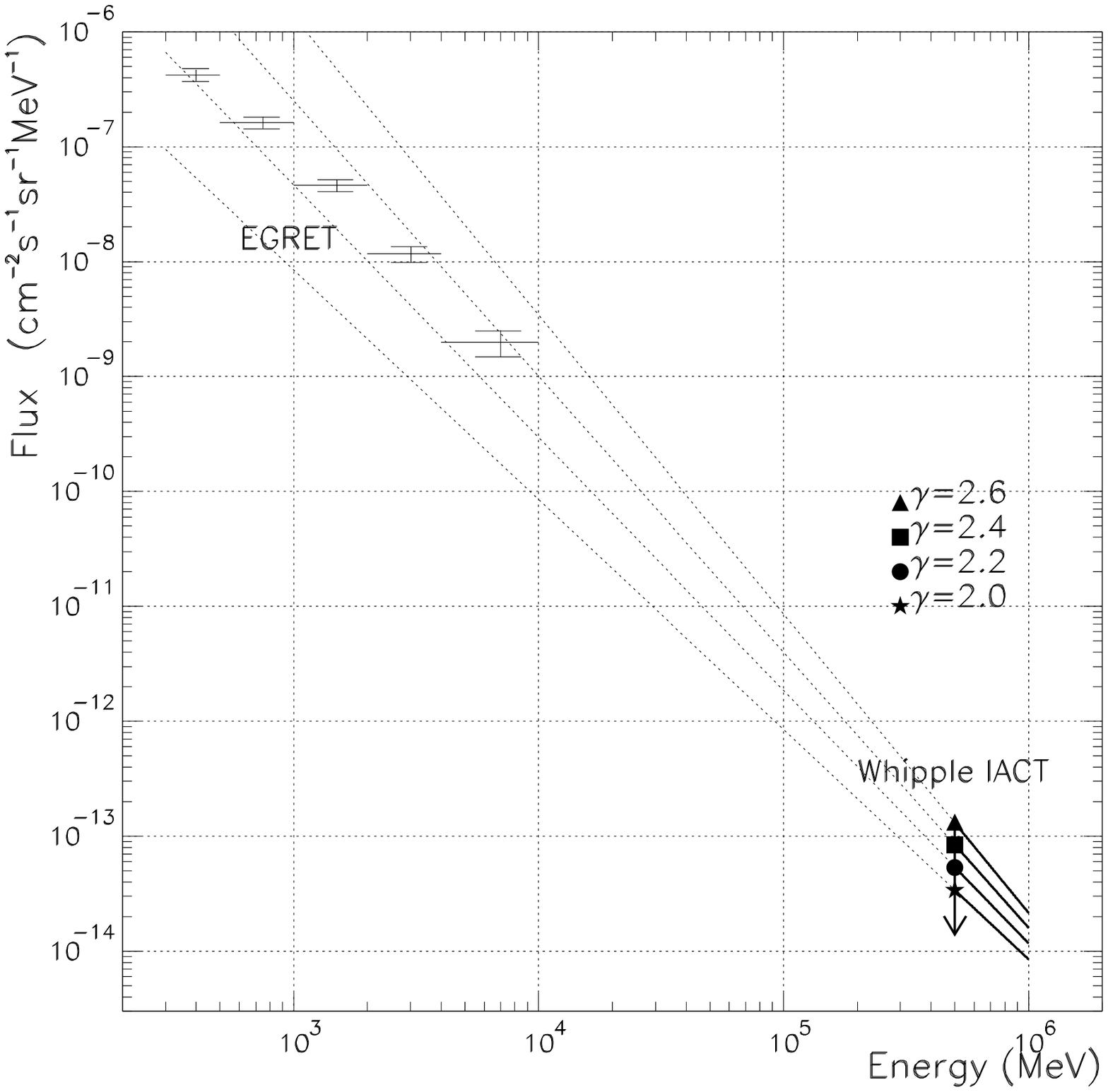}}
\end{center}
 \figcaption[m42d] {The EGRET differential spectrum of the diffuse
emission observed in the portion of sky defined by {$\rm
38.5^o<l<41.5^o$} and {$\rm 2^o<b<2^o$} is compared to the upper
limits obtained above 500~GeV in 1999 with the Whipple Telescope. 
The EGRET data points in the upper left of the figure with the $1\sigma$
error bars on fluxare shown.  The 99.9 \% CL Whipple upper limits are shown in the lower
right of the figure for assumed values of 2.0, 2.2, 2.4 and 2.6 (star,
circle, square and triangle) for
the differential spectral index $\gamma$.  If these upper limits are
extrapolated back to EGRET energies assuming a simple power law
spectrum, the four dotted lines shown are obtained.  It is then
apparent that, if $\gamma$ were less than about 2.2, a TeV signal
should have been detected. The various dotted lines show a power law 
extrapolation down to the EGRET regime of our upper limits. Each line 
corresponds to a different assumption for the differential spectral 
index. The upper limit obtained with the smallest spectral index is not 
compatible with the flux recorded by EGRET above 1~GeV.
 \label{fig7}}

\clearpage
\begin{center}
\mbox{\epsfysize=0.60\textheight\epsfbox{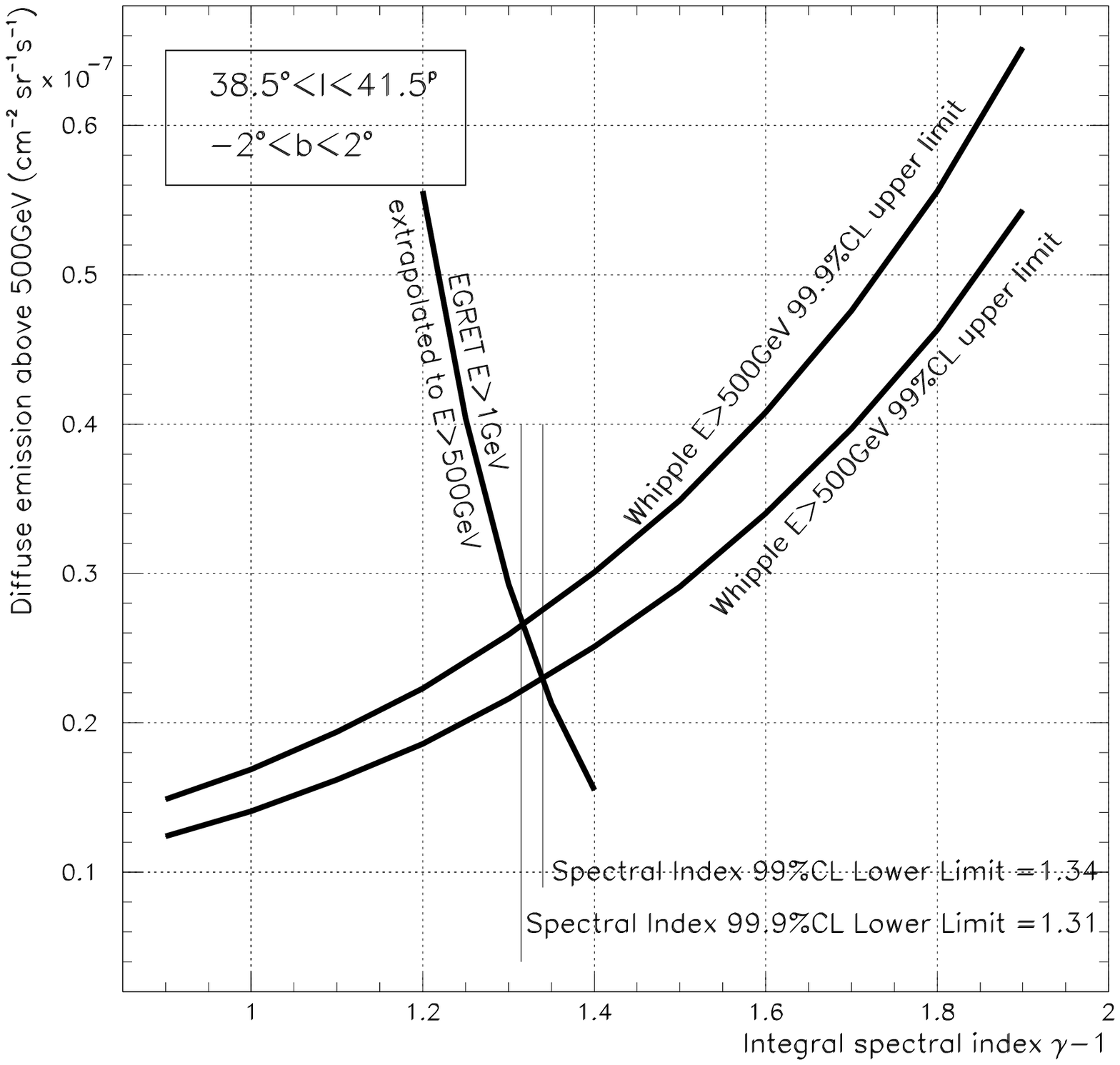}}
\end{center}
 \figcaption[m42d]
{Diffuse emission upper limits derived from 1999 observations
represented as a function of the assumed integral spectral index and 
compared with the flux measured by EGRET above 1~GeV in the same
field and extrapolated to 500~GeV with various spectral indices. This
comparison yields a lower limit ($99.9\%$ confidence level) for the
integral spectral index of 2.31.   
 \label{fig8}}

\end{document}